\documentclass[aps,prb,twocolumn,epsfig,superscriptaddress,floatfix]{revtex4}
\makeatletter
\def\thepage{\@arabic\c@page}
\def\@pnumwidth{2em}

\makeatother
\usepackage[dvips]{graphicx}
\usepackage{epsfig}
\usepackage{amssymb}

\newcounter{Table}
\newenvironment{Table}[1]{\refstepcounter{Table}
   \begin{center} Table \theTable. \\  \rm #1 \\[2ex]
  }{\end{center}}

\begin{document}

\title{
Isotropic-nematic transition in liquid crystals confined between rough
walls
}

\author{D. L. Cheung}
\affiliation{Department of Physics and Centre for Scientific Computing,\\
University of Warwick, Coventry, CV4 7AL, UK}
\email{david.cheung@warwick.ac.uk}
\author{F. Schmid}
\affiliation{Fakult\"{a}t f\"{u}r Physik, Universit\"{a}t,
 Bielefeld, 33615 Bielefeld, Germany}

\begin{abstract}

The effect of rough walls on the phase behaviour of a confined liquid
crystal (LC) fluid is studied using constant pressure Monte Carlo
simulations. The LC is modelled as a fluid of soft ellipsoidal
molecules and the rough walls are represented as a hard wall with a
number of molecules randomly embedded in them. It is found that the
isotropic-nematic (IN) transition is shifted to higher pressures for
rougher walls.
\end{abstract}
\maketitle

\newpage

\section{Introduction}

The phase behaviour of many fluids alters when they are 
confined\cite{gelb1999a}. This is particularly true for ordered
fluids such as liquid crystals (LC)\cite{jerome1992a}. As well as being
interesting in its own right, the effect of confinement LCs is of
technological significance. Many applications of LCs
depend crucially on this interaction.

Because of this interest there have been many studies of confined
LCs. Simulations have studied many aspects of the effect
of confinement on LCs, such as wetting behaviour\cite{dijkstra2001a}
 and anchoring strength\cite{allen2002a}. The
effect of confinement on the phase behaviour has also been studied\cite{quintana2002a,steuer2004a}.
In a recent paper the effect of rough walls on the behaviour of
confined LCs was examined\cite{cheung2005a}. Specifically the structure of the fluid and the surface
anchoring was studied. It was found that the surface anchoring became
weaker as the surfaces became rougher, which may have implications for
the phase behaviour of the confined fluid. The aim of the present
paper is to study the effect of rough surfaces on the
isotropic-nematic (IN) transition of a confined LC.

\section{Model and simulation details}

In order to simulate both a large number of fluid molecules for a
reasonable range of pressures and wall roughnesses simple models for both the molecule-molecule and molecule-wall 
interactions are used.

A  simplified version of the Gay-Berne (GB)
potentia\cite{gay1981a}l is employed. The interaction between two
molecules $i$ and $j$, with positions $\mathbf{r}_i$ and
$\mathbf{r}_j$, and orientations $\mathbf{u}_i$ and $\mathbf{u}_j$ is given by
\begin{equation}\label{eqn:vse1}
  V(\mathbf{r}_{ij},\mathbf{u}_i,\mathbf{u}_j)= 
  \left\{ \begin{array}{c @{,\:\:} c}
  4\epsilon_0\left[\rho^{-12}-
    \rho^{-6}\right]+\epsilon_0
  &
  \rho \le 2^{1/6}
  \\
  0
  &
  \mbox{otherwise}
  \end{array}
  \right.
\end{equation}
where $\mathbf{r}_{ij}=\mathbf{r}_i-\mathbf{r}_j$ and
\begin{equation}
  \rho(\mathbf{r}_{ij},\mathbf{u}_i,\mathbf{u}_j)=
  \frac{r_{ij}-\sigma(\hat{\mathbf{r}}_{ij},\mathbf{u}_i,\mathbf{u}_j)+\sigma_0}
  {\sigma_0}.
\end{equation}
$r_{ij}=|\mathbf{r}_{ij}|$, and 
$\hat{\mathbf{r}}_{ij}=\mathbf{r}_{ij}/r_{ij}$. $\sigma_0$ is the
molecular width, which defines the length scale used throughout this
paper. $\sigma(\hat{\mathbf{r}}_{ij},\mathbf{u}_i,\mathbf{u}_j)$ is
the shape function given by\cite{berne1975a}
\begin{eqnarray}
  \sigma(\hat{\mathbf{r}}_{ij},\mathbf{u}_i,\mathbf{u}_j) &=&
  \sigma_0
  \left\{1-\frac{\chi}{2}\left[
    \frac{(\hat{\mathbf{r}}_{ij}.\mathbf{u}_i+
           \hat{\mathbf{r}}_{ij}.\mathbf{u}_j)^2}
         {1+\chi\mathbf{u}_i.\mathbf{u}_j}\right.\right.
         \nonumber\\
        & &+\left.\left.
    \frac{(\hat{\mathbf{r}}_{ij}.\mathbf{u}_i-
           \hat{\mathbf{r}}_{ij}.\mathbf{u}_j)^2}
         {1-\chi\mathbf{u}_i.\mathbf{u}_j}\right]\right\}^{-1/2}
     \label{eqn:vse3}
\end{eqnarray}
where $\chi=(\kappa^2-1)/(\kappa^2+1)$, with the elongation $\kappa=3$. 
This approximates the contact distance between two
ellipsoids. $\epsilon_0$ and $\sigma_0$ define the energy and length
scales used throughout this paper, with other quantities given in
terms of these. Specifically the reduced pressure is
$P^*=P\sigma_0^3/\epsilon_0$ and the reduced temperature is
$T^*=k_BT/\epsilon_0$. 
As the potential is fully repulsive,
temperature is not a significant variable and is set $T^*=0.5$ throughout.

The walls are represented by a hard core potential acting on the centres
of mass of the fluid molecules, giving rise to homeotropic anchoring. Roughness is introduced by 
embedding some molecules in the wall\cite{cheung2005a}. These
are given random positions in the $xy$ plane and random orientations
which are held fixed during the simulations. It should be noted that
these are used as a convenient method to introduce
disorder into the surface-molecule interaction and do
not correspond to real molecules. The roughness of the wall is
characterised by the density of these embedded molecules $\Sigma$.

As the aim of this study is to investigate the phase behaviour, constant
$NPT$ simulations are more appropriate than constant $NVT$
simulations\cite{allentildesley}. To retain a constant $\Sigma$, volume moves involve only changes in the
$z$ box length. The phase behaviour is probed using a series of
simulations at different pressures, starting from a high pressure
$P^*=2.50$ in the nematic phase. The simulated systems comprised
1200 fluid molecules and up to 60 molecules in the walls. The
fixed area of each wall was 154.9 $\sigma_0^2$. The cell width was in the range 22 - 28 $\sigma_0$. To ensure some
sampling of surface configurations three different walls
were studied for each value of $\Sigma$. The order parameters and
directors for these are presented in Tab. \ref{tab:wall}.

\begin{Table}{\label{tab:wall}Order parameter and directors for the
    wall configurations studied.}
\begin{tabular}{c c c c}
\hline
$\Sigma$  &     &    $S_2^{wall}$ &  $ \mathbf{n}^{wall}$     \\
\hline
   0.1    &  1  &      0.165    &   ( 0.261,-0.560, 0.786)  \\
          &  2  &      0.181    &   (-0.520, 0.404,-0.753)  \\
          &  3  &      0.145    &   (-0.392,-0.919, 0.035)  \\
   0.2    &  1  &      0.153    &   ( 0.013, 0.054, 0.952)  \\
          &  2  &      0.304    &   ( 0.638, 0.661, 0.396)  \\
          &  3  &      0.235    &   (-0.338,-0.937, 0.089)  \\         
   0.3    &  1  &      0.217    &   ( 0.051,-0.177, 0.983)  \\
          &  2  &      0.220    &   ( 0.963, 0.016, 0.268)  \\
          &  3  &      0.160    &   ( 0.009, 0.992, 0.128)  \\
   0.4    &  1  &      0.329    &   (-0.745, 0.035, 0.666)  \\
          &  2  &      0.382    &   (-0.884, 0.539, 0.121)  \\
          &  3  &      0.262    &   (-0.062,-0.329, 0.942)  \\         
\hline
\end{tabular}
\end{Table}   

Due to the number of different
systems considered (3 for each non-zero $\Sigma$ and one smooth wall) only a coarse sampling of pressure is attempted; the pressure
being changed in steps of 0.02 (in reduced units) near the NI
transition. This limits the accuracy at which transition
pressures may be determined. While techniques such as
Gibbs Ensemble Monte Carlo (GEMC)\cite{panagiotopoulos1987a,panagiotopoulos1987b} or grand 
canonical Monte Carlo (GCMC)\cite{allentildesley} simulations may 
be used to investigate phase equilibria the densities involved lead to 
unacceptably low acceptance rates for particle transfers. Studying
more elongated molecules \cite{allen1995a,schilling2005b} (with lower $\rho_{NI}$) or softer interaction
models\cite{schilling2005a} would allow the use of these techniques.

As the nematic phase has orientational order which is lacking in the
isotropic phase, the phase of the system may be determined by
calculating the orientational order parameter. This is given by the
largest eigenvalue of the ordering tensor
\begin{equation}
  Q_{\alpha\beta}=\sum_{i=1}^{N}\left(
    \frac{3}{2}u_{i\alpha}u_{i\beta}-\frac{1}{2}\delta_{\alpha\beta}
    \right)
\end{equation}
where $\delta_{\alpha\beta}$ is the Kronecker delta function. The
 order parameter $S_2$ is then determined by diagonalising this.
For a nematic $S_2$ is
typically $>0.4$, while lower values are typical of an isotropic state.

As the presence of surfaces gives rise to an inhomogeneous distribution
of molecules it is useful to consider quantities such as the density
or order parameter in different regions of the simulation cell. 
As previous studies indicate that a layer of well ordered molecules
forms even for rough walls\cite{cheung2005a}, the order parameter in the central half of
the cell $S_2^{bulk}$ ($l_z/4$ to $3l_z/4$) may be taken to be the most useful variable for
characterising the phase of the confined fluid
 The variation of
density and order may be examined more closely by the density and
order parameter profiles.

\section{Results and Discussion}

\subsection{Phase behaviour}

\begin{figure}
\includegraphics[width=4.2cm]{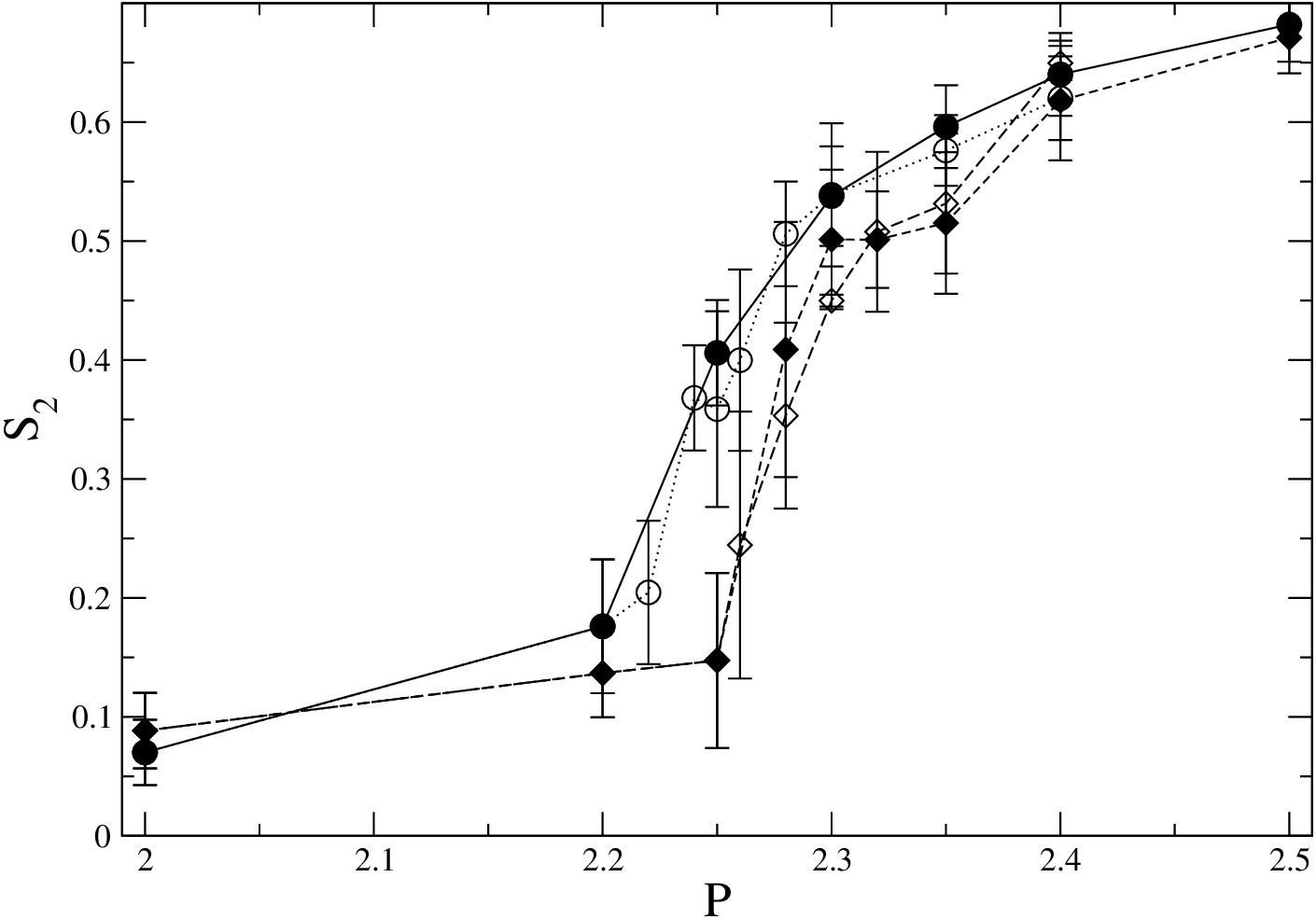}
\includegraphics[width=4.2cm]{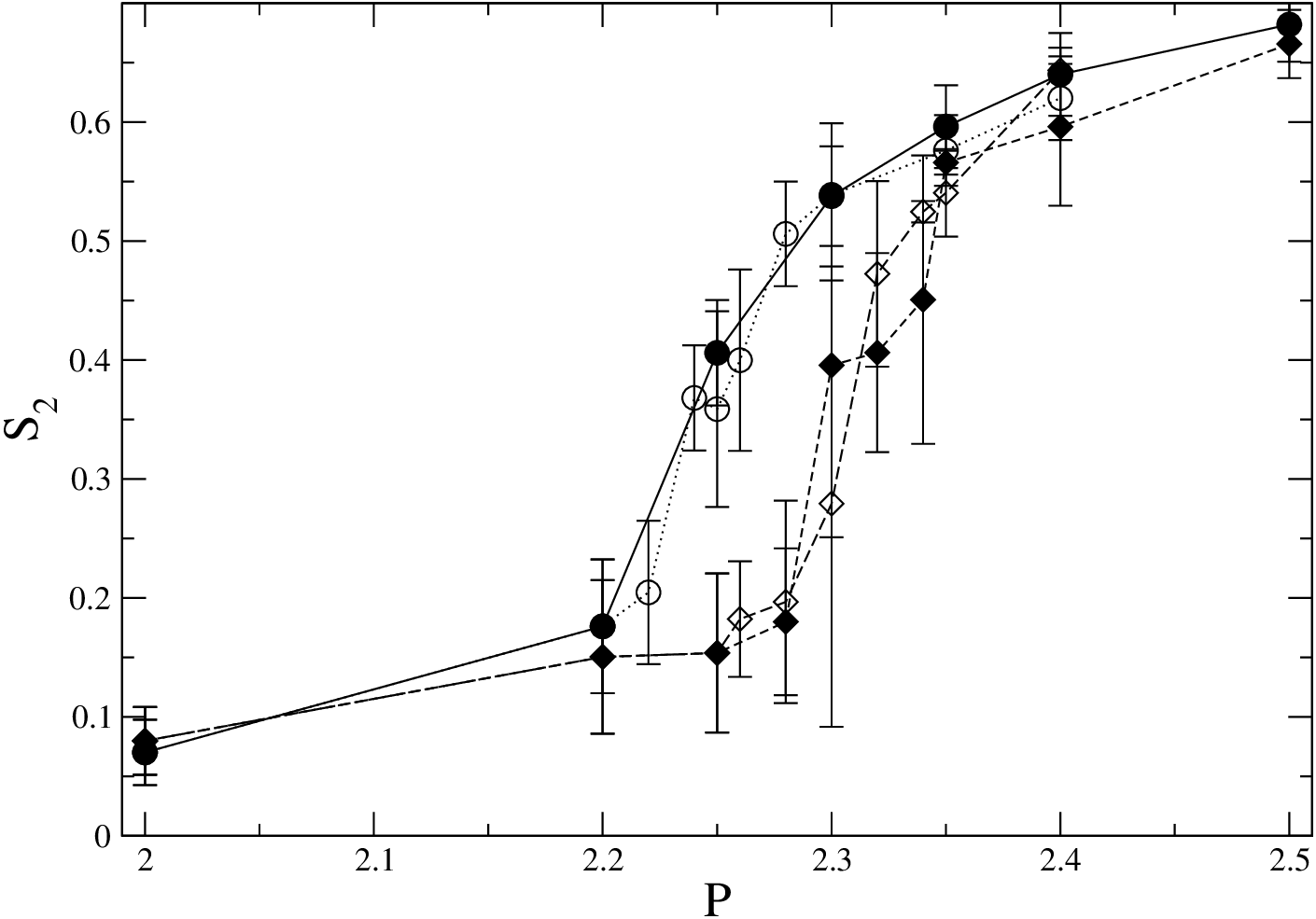}
\includegraphics[width=4.2cm]{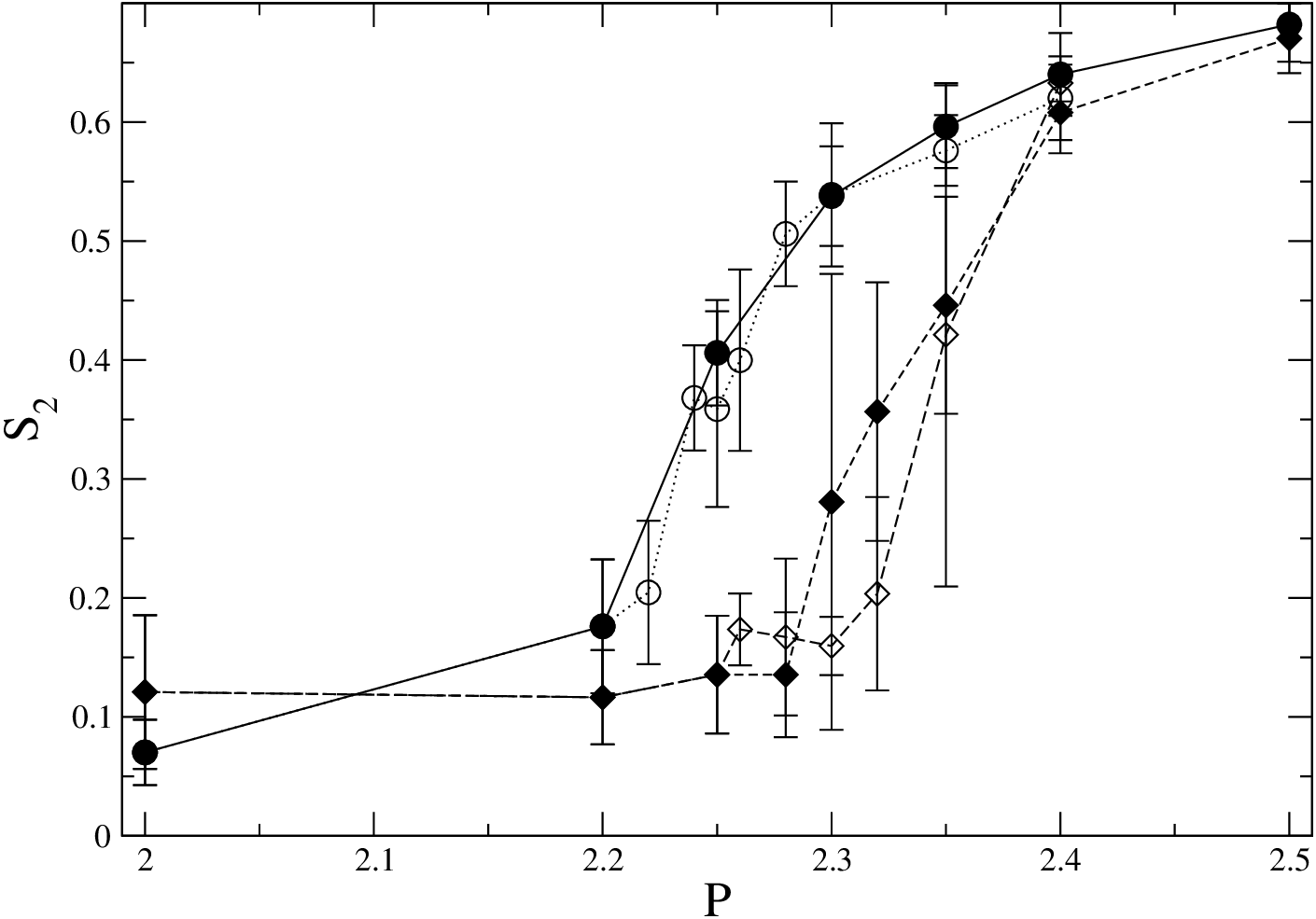}
\includegraphics[width=4.2cm]{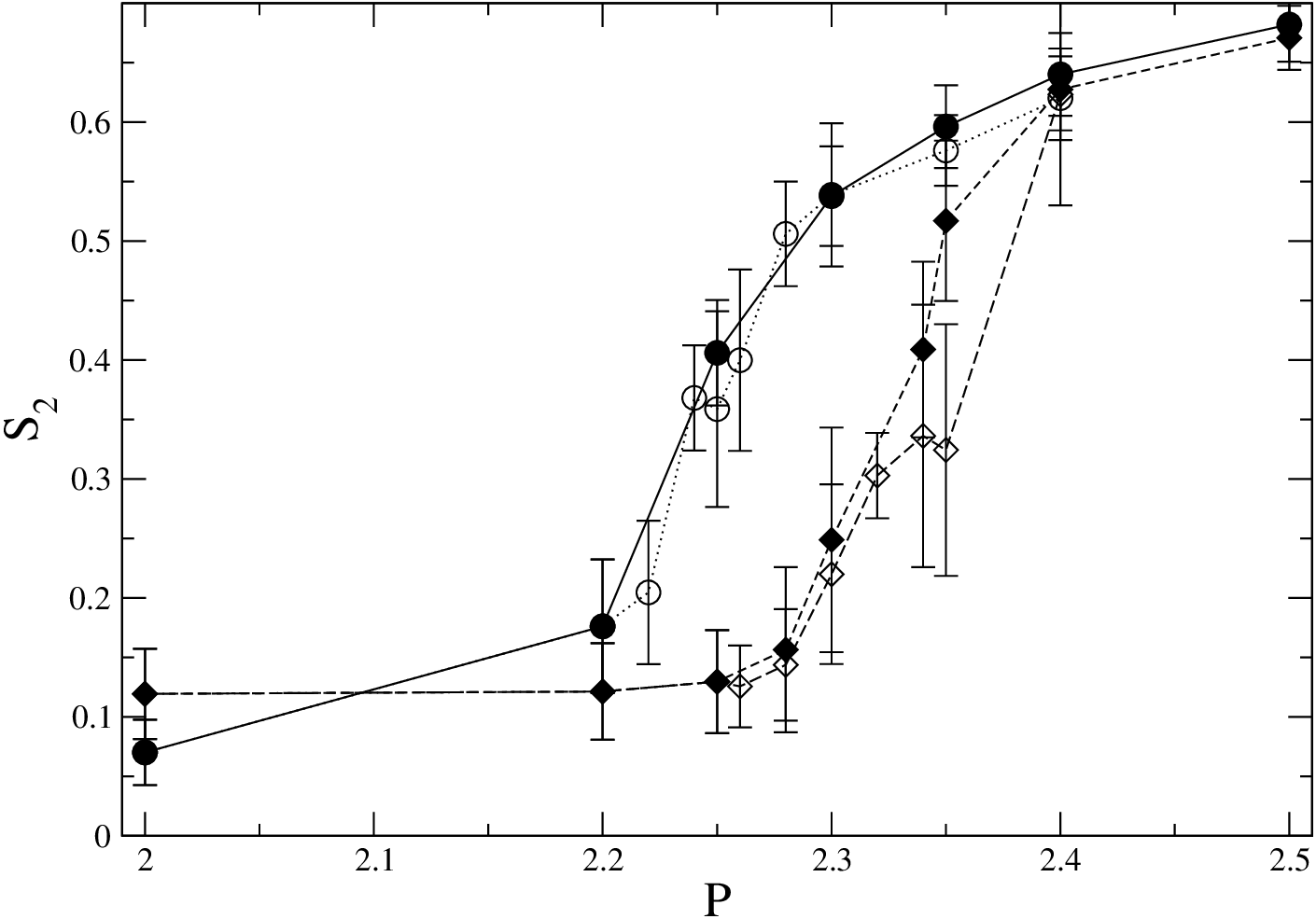}
\caption{\label{fig:phase}
Variation of bulk order parameter with pressure. Solid line with
filled circles shows $\Sigma=0.0$ (expansion), dotted line with open
circles shows $\Sigma=0.0$, dashed line with filled diamonds shows (a)
$\Sigma=0.1$, (b) $\Sigma=0.2$, (c) $\Sigma=0.3$, and (d) $\Sigma=0.4$
(expansion) and the long dashed line with open diamonds shows (a) 
$\Sigma=0.1$, (b) $\Sigma=0.2$, (c) $\Sigma=0.3$, and (d) $\Sigma=0.4$ 
(compression).}
\end{figure}

Shown in Fig. \ref{fig:phase} $S_2^{bulk}$ as a function of $P$. As
can be seen, on expansion $S_2^{bulk}$ gradually decreases until $P^*=2.25$ ($S_2^{bulk} \approx 0.4$), before falling
to an isotropic state at $P^*=2.20$. For this system the bulk
isotropic-nematic transition occurs at $P_{NIbulk}^*=2.30$\cite{schmid2002b}.
Note that due to finite size effects the order parameter does not go
exactly to 0\cite{frenkel1984a}. The system shows a similar transition
on compression for an isotropic starting state.  The
lowering of $P_{NI}$ relative to the
bulk phase is caused by the formation of a highly ordered layer near
the walls\cite{cleaver1997a,steuer2004a}.

This behaviour is similar to the rough wall with
$\Sigma=0.1$(Fig. \ref{fig:phase}a) . However, the NI transition seems to be shifted to a higher pressure (between $P^*=2.25$
and $P^*=2.30$). Away from the transition the behaviour of
$S_2^{bulk}$ is similar to that of the smooth wall system. For pressures above the $P_{BulkNI}$ the fluid in
the cell bulk would be nematic anyway so the rough walls would have
little effect on $S^{bulk}$ in this case, while for $P$ much less than 
$P_{NI}$ the anchoring effect of the walls is
insufficient to lead to ordering in the cell bulk.

The effect of further increasing the surface roughness may be seen in
Fig. \ref{fig:phase}b, which shows $S_2^{bulk}$ against $P^*$ for the
$\Sigma=0.2$ wall. As may
be seen the NI transition is shifted to a high pressure compared to the
$\Sigma=0.1$ wall ($P_{NI}^*\approx2.30$).  Figure \ref{fig:phase}(c) and
(d) show the phase diagram for walls with $\Sigma=0.3$ and
$\Sigma=0.4$ respectively. While $P_{NI}$ increases on going from
$\Sigma=0.2$ to $\Sigma=0.3$, there is a much smaller difference
between $\Sigma=0.3$ and $\Sigma=0.4$, suggesting that the effect of
the rough walls becomes saturated with increasing roughness, similar
to the anchoring coefficient\cite{cheung2005a}. 
 The change in order 
parameter through the NI transition also appears to become smoother
as $\Sigma$ increases. This may be due to averaging
over different wall configurations as these
may have different $P_{NI}$. This change in the location of the
IN transition may also be seen from
the large error bars for $S_2^{bulk}$ for pressures near $P_{NI}$.

The approximate transition pressures and order parameters are presented in
Tab. \ref{tab:ni_trans}. These generally increase with surface roughness. This may be
attributed to the rough walls disturbing the highly ordered surface
layer, which disrupts the order in the bulk. This can
also be seen from the decrease in the surface anchoring strength as
$\Sigma$ increases\cite{cheung2005a}.

\begin{Table}{\label{tab:ni_trans}Approximate transition pressures and
  order parameters. Errors in the final decimal place are in
parenthesises}
\begin{tabular}{c c c c c c c c c}
\hline
$\Sigma$  &  $P_N^{exp}$  &  $P_I^{exp}$  &  $P_N^{cmp}$  &
$P_I^{cmp}$ & $S_2^{Nexp}$ &  $S_2^{Iexp}$  &  $S_2^{Ncmp}$  & $S_2^{Icmp}$ \\
\hline
 0.0  &  2.25 & 2.20 & 2.26 & 2.22  & 0.41(4) & 0.18(6) & 0.40(7) & 0.20(6) \\
 0.1  &  2.30 & 2.25 & 2.30 & 2.25  & 0.50(6) & 0.15(7) & 0.45(1) & 0.15(7) \\
 0.2  &  2.35 & 2.25 & 2.34 & 2.30  & 0.57(1) & 0.15(7) & 0.52(9) & 0.20(8) \\
 0.3  &  2.40 & 2.28 & 2.40 & 2.30  & 0.61(3) & 0.14(5) & 0.63(2) & 0.16(2) \\
 0.4  &  2.35 & 2.28 & 2.40 & 2.28  & 0.52(7) & 0.16(7) & 0.62(9) & 0.14(5) \\
\hline
\end{tabular}
\end{Table}

\subsection{Fluid Structure}

\begin{figure}
\includegraphics[width=4.2cm]{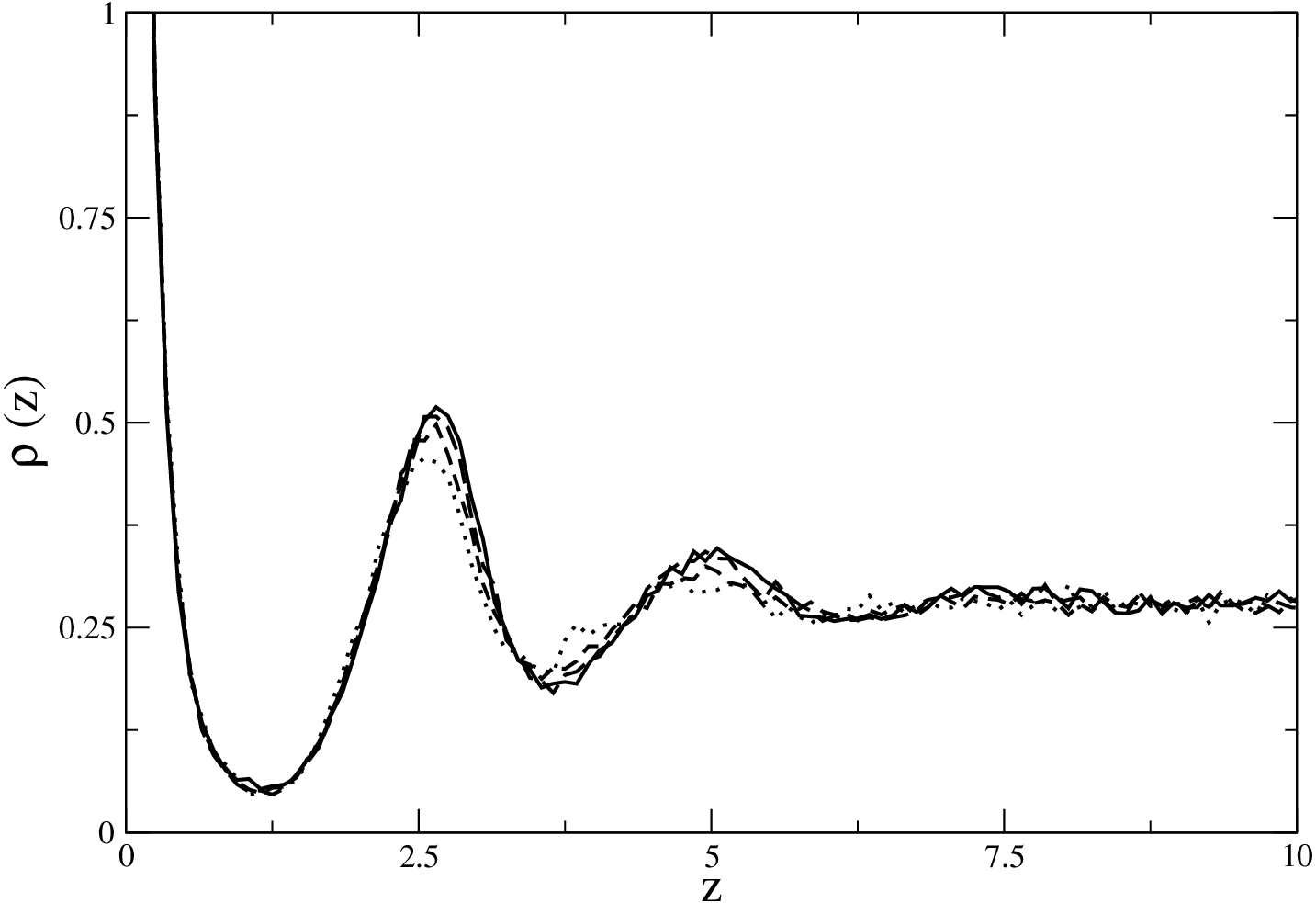}
\includegraphics[width=4.2cm]{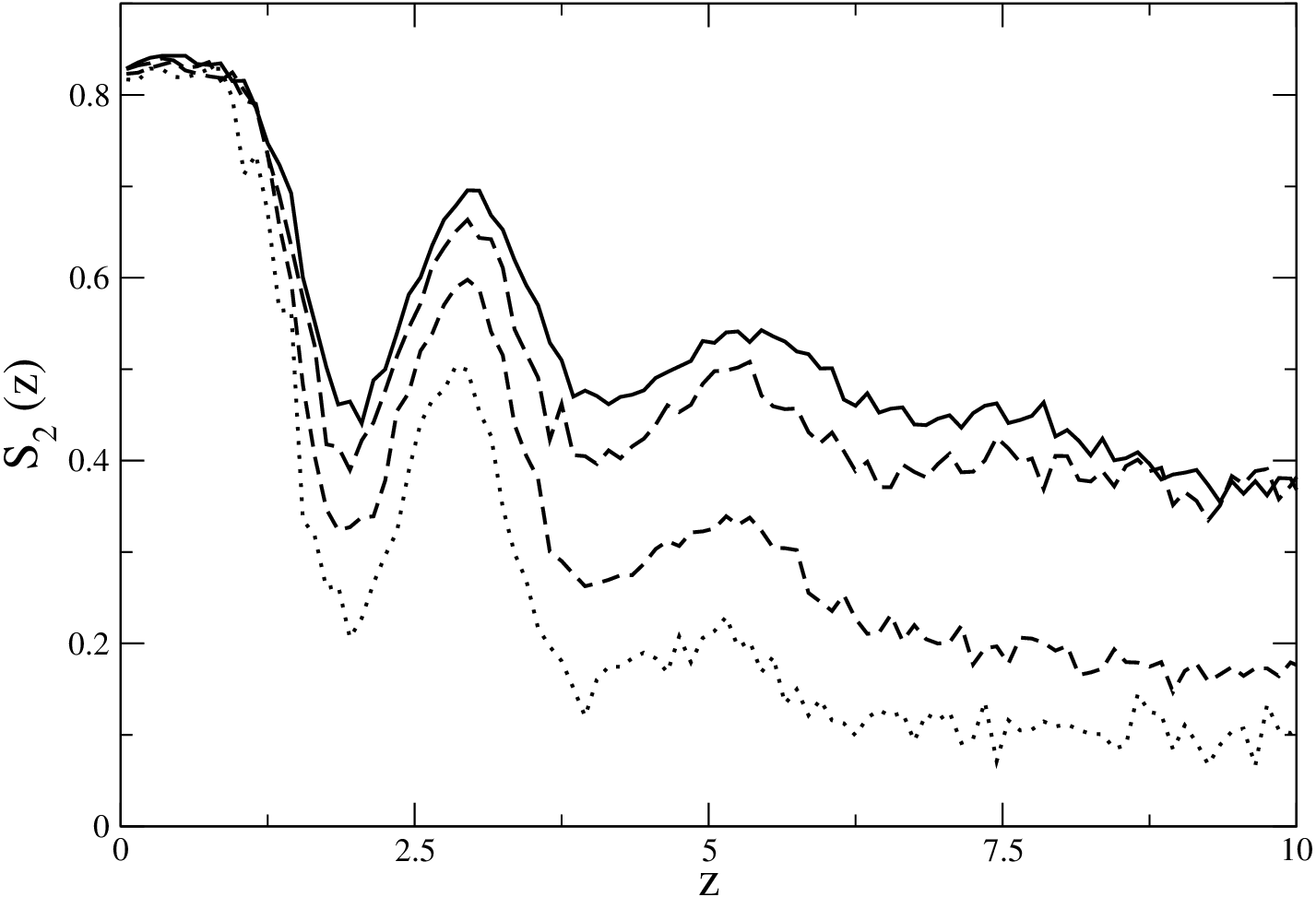}
\caption{\label{fig:prof_sigma0.0}
(a) Density profiles for the smooth wall system. Solid line is
$P^*=2.25$ (expansion), dotted line is $P^*=2.20$ (expansion), dashed
line is $P^*=2.22$ (compression), and dot-dashed line is $P^*=2.26$ 
(compression) (b) Order parameter profiles for smooth wall
system. Symbols as in (a).}
\end{figure}

Shown in Fig. \ref{fig:prof_sigma0.0}a are density profiles for the smooth
wall system. These are shown for pressures just above and below the 
isotropic-nematic transition for both expansion and compression (for
the expansion runs these are $P_{nem}=2.25$ and $P_{iso}=2.20$ and for
the compression runs $P_{nem}=2.26$ and $P_{iso}=2.22$). As
can be seen there is a noticeable decrease in the height of the density
peak at around $z \approx 2.5$ when going from the nematic to isotropic phases 
($P^*=2.25 \rightarrow 2.20$). However, on going from the isotropic to nematic
phase ($P^*=2.22 \rightarrow 2.26$) there is a much smaller change in
the density profile.

The order parameter profiles for these pressures are shown in Fig. 
\ref{fig:prof_sigma0.0}b. For all pressures there is a well
ordered layer near the wall, which has similar values for all
pressures. More variation is seen around the second peak
$z\approx2.5$. For the isotropic fluids the order parameter profile is
typically less than 0.1, while for the nematic $S_2^{bulk}(z)\approx
0.4$. 

\begin{figure}
\includegraphics[width=4.2cm]{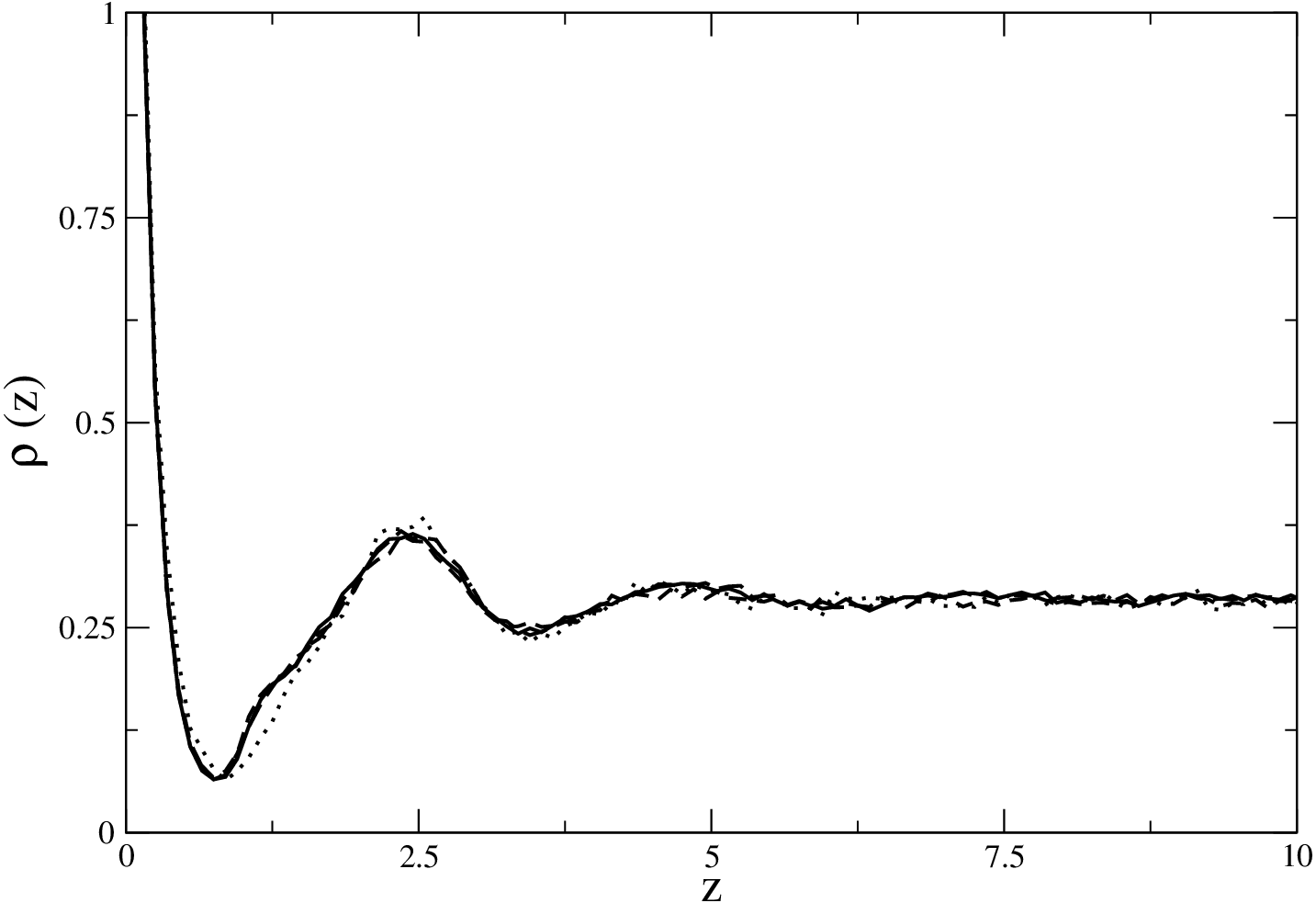}
\includegraphics[width=4.2cm]{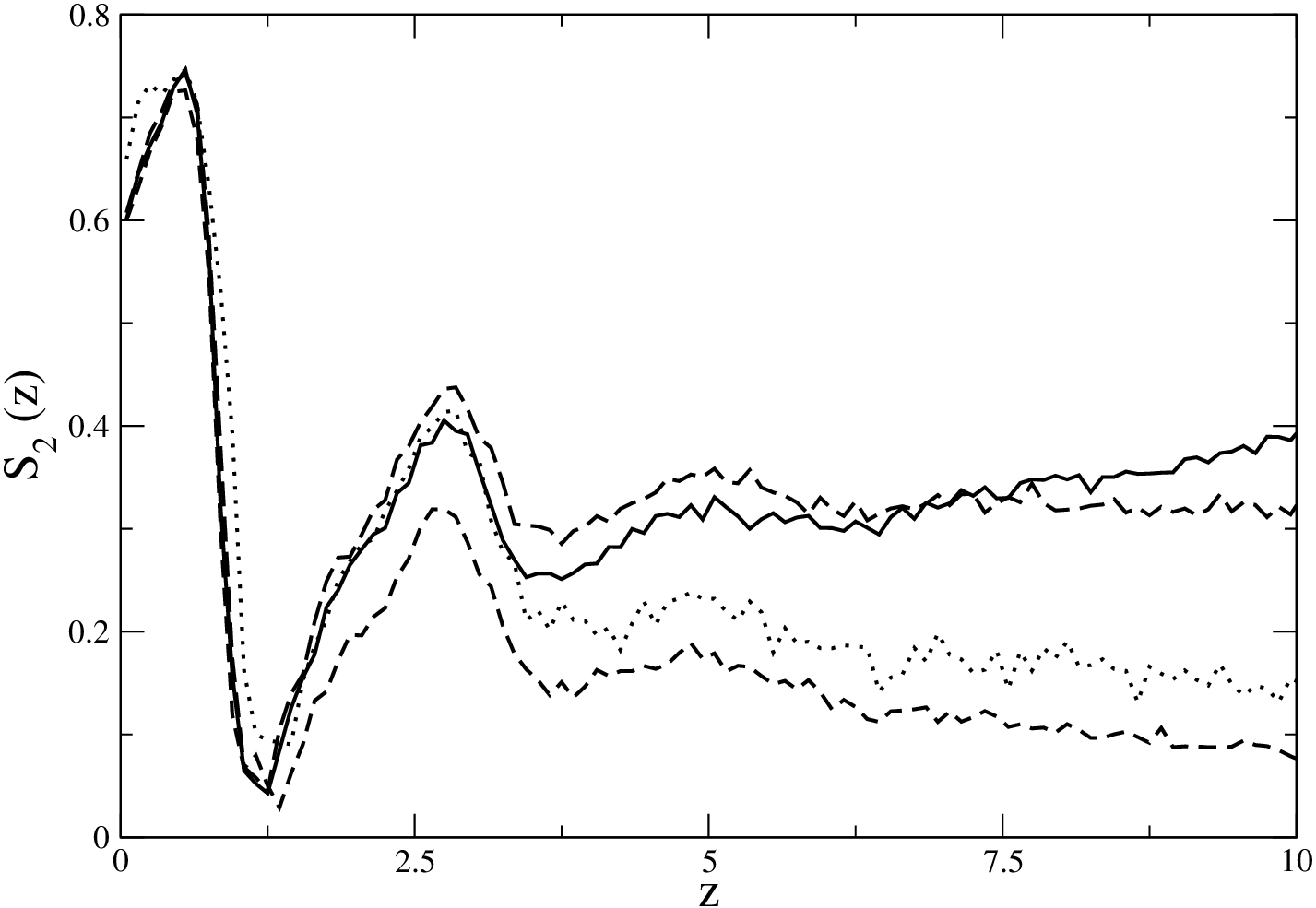}
\caption{\label{fig:prof_sigma0.2}
(a) Density profiles for rough wall with $\Sigma=0.2$. Solid line is
$P^*=2.30$ (expansion), dotted line is $P^*=2.28$ (expansion), dashed
line is $P^*=2.28$ (compression), and dot-dashed line is $P^*=2.32$ 
(compression) (b) Order parameter profiles for rough wall with $\Sigma=0.2$.
Symbols as in (a).}
\end{figure}

Density and order parameter profiles for the rough wall with
$\Sigma=0.2$ are shown in Fig. \ref{fig:prof_sigma0.2}. The expansion runs are given at
$P_{nem}=2.30$ and $P_{iso}=2.28$ and for the compression runs
$P_{nem}=2.28$ and $P_{iso}=2.32$. Compared to
the smooth wall the density profiles are less structured, with only a
single peak at $z\approx2.5$ visible for the lowest pressures
shown, which  also appears wider than the equivalent peak for the
smooth wall, due to the disruption of the
layer structure due to the surface disorder\cite{dong1999a}.

The effect of the rough walls is more prominent for the
order parameter profiles (fig. \ref{fig:prof_sigma0.2}b). Near the wall 
$S_2^{bulk}(z)$is significantly lower than for the smooth wall, due to
the disordering effect of the wall molecules. The
order minima is deeper (the minimum value$\approx 0.1$ for all 
pressures shown) and is moved closer to the wall
($z\approx1.1$). However, this does
not extend into the bulk fluid.

\begin{figure}
\includegraphics[width=4.2cm]{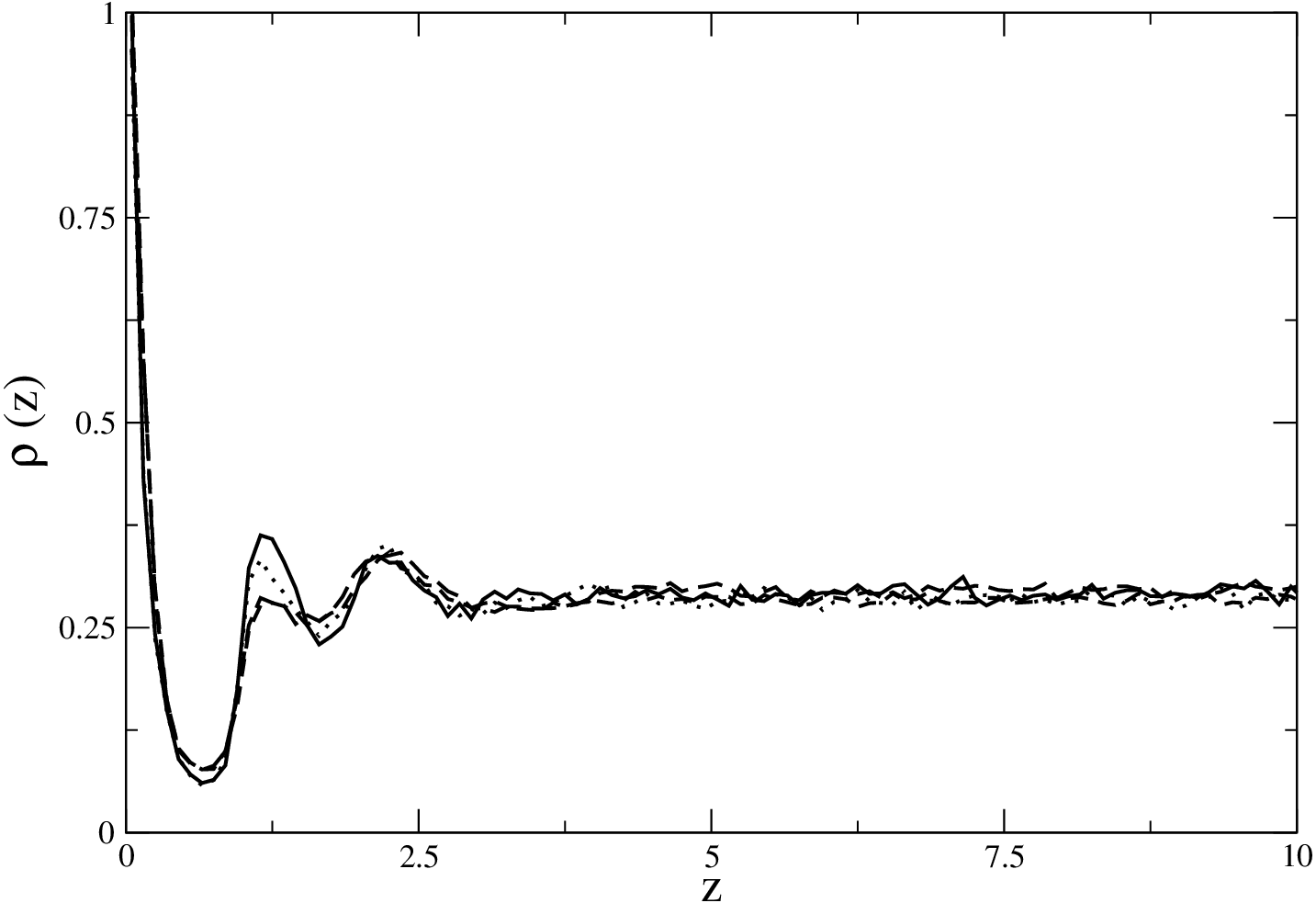}
\includegraphics[width=4.2cm]{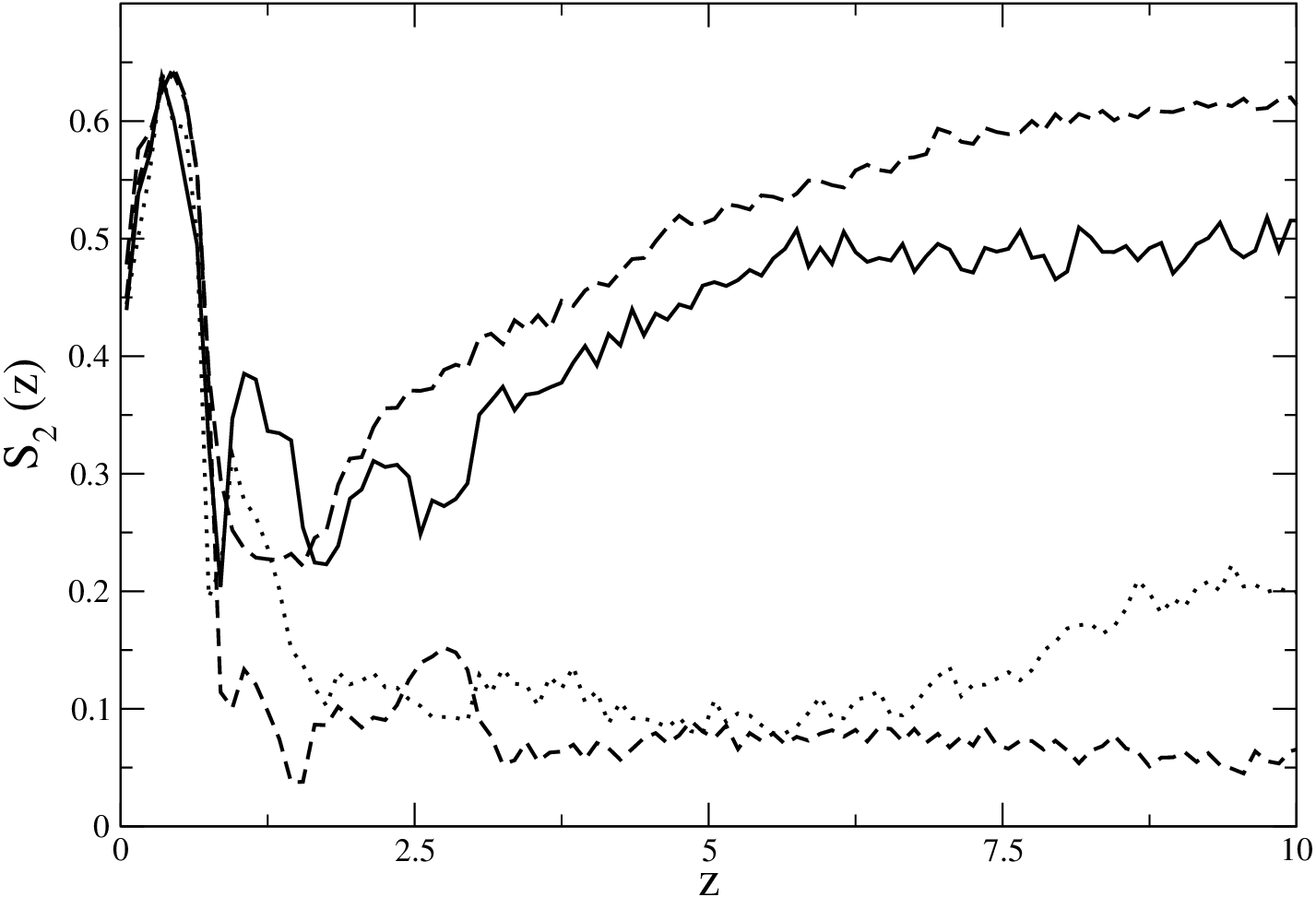}
\caption{\label{fig:prof_sigma0.4}
(a) Density profiles for rough wall with $\Sigma=0.4$. Solid line is
$P^*=2.35$ (expansion), dotted line is $P^*=2.28$ (expansion), dashed
line is $P^*=2.28$ (compression), and dot-dashed line is $P^*=2.40$ 
(compression) (b) Order parameter profiles for rough wall with $\Sigma=0.2$.
Symbols as in (a).}
\end{figure}

Density and order parameter profiles for $\Sigma=0.4$ are shown in
Fig. \ref{fig:prof_sigma0.4}. Profiles for the expansion runs are
at $P_{nem}^*=2.35$ and $P_{iso}^*=2.28$ and for the compression runs
$P_{nem}^*=2.28$ and $P_{iso}^*=2.40$. The density
profile shows 
even less structure than for $\Sigma=0.2$, with only a density peak at 
$z\approx2.5$ visible. However, this peak has a broad shoulder starting at
$z\approx2.0$, corresponding to molecules lying in the plane of
the wall in the region between the two layers\cite{cheung2005a}. This
is similar to the behaviour seen in simulations of smectic LCs\cite{frenkel1995a}. The effect of this bimodal distribution
may be seen by the depletion of the order parameter in this region and
by a peak in the biaxiality profile (c.f. Fig 2d of Ref. \cite{cheung2005a}). 

\section{Conclusions}

The phase behaviour of a model liquid crystal confined between two
rough walls has been studied. Particular attention is paid to its
behaviour in the vicinity of the isotropic-nematic transition. 

For smooth walls it is found that the nematic-isotropic transition is
shifted to lower pressures compared to the bulk. This may be
attributed to the anchoring effect of the walls. As the walls become
rougher, this effect is weakened so the transition is shifted to
higher pressures. For pressures away from the transition pressure the
rough walls appear to have a smaller effect.


On examining the structure of the confined fluid, it is found
that the density distribution changes little with pressure, rather
changes in this are largely due to increasing wall roughness. The
behaviour of the order parameter near the wall is likewise determined
by the wall roughness.

To summarize, we find that surface roughness not only decreases the
nematic order in the vicinity of the surface, but also shifts the
nematic-isotropic transition to higher pressures in thin slabs. It
would be interesting to study other quantities, in particular, the 
anchoring direction. On flat surfaces, the anchoring
direction is homeotropic. On rough surfaces, we observe that the system
sometimes switches to planar anchoring. However, this depends
sensitively on the particular realization of the disorder, and much
more extensive simulations of much larger systems would be necessary
to clarify this issue.

\section*{Acknowledgements}

Computational resources for this work were provided by the Centre for 
Scientific Computing of the University of Warwick. The authors wish to
acknowledge helpful conversions with Michael Allen and Tanja
Schilling. This work was funded by the UK EPSRC.

\end{document}